\documentclass{mpe_report}

\usepackage{psfig,graphicx,epsfig}
\usepackage{color}
\usepackage{amsmath,amssymb,epic,eepic,array}

\begin{document}

\pagenumbering{arabic}
\setcounter{page}{108}

\renewcommand{\FirstPageOfPaper }{108}\renewcommand{\LastPageOfPaper }{111}

\def\aap{A\&A}
\def\apj{ApJ}
\def\apjl{ApJL}
\def\araa{ARA\&A}
\def\mnras{MNRAS}
\def\nat{Nature}
\def\apss{Ap\&SS}

\title{Polarisation of high-energy emission in a pulsar striped wind}
\author{J. P\'etri\inst{1} \and J. Kirk}
\institute{Max-Planck-Institut f\"ur Kernphysik,
  Saupfercheckweg 1, 69117 Heidelberg - Germany}
\maketitle

\begin{abstract}
  Recent observations of the polarisation of the optical pulses from
  the Crab pulsar motivated detailed comparative studies of the
  emission predicted by the polar cap, the outer gap and the two-pole
  caustics models.
  
  In this work, we study the polarisation properties of the
  synchrotron emission emanating from the striped wind model. We use
  an explicit asymptotic solution for the large-scale field structure
  related to the oblique split monopole and valid for the case of an
  ultra-relativistic plasma.  This is combined with a crude model for
  the emissivity of the striped wind and of the magnetic field within
  the dissipating stripes themselves. We calculate the polarisation
  properties of the high-energy pulsed emission and compare our
  results with optical observations of the Crab pulsar.  The resulting
  radiation is linearly polarised. In the off-pulse region, the
  electric vector lies in the direction of the projection on the sky
  of the rotation axis of the pulsar, in good agreement with the data.
  Other properties such as a reduced degree of polarisation and a
  characteristic sweep of the polarisation angle within the pulses are
  also reproduced.
\end{abstract}

\section{Introduction}

The high-energy, pulsed emission from rotating magnetised neutron
stars is usually explained in the framework of either the polar cap or
the outer gap models.  Although the existence of such gaps is
plausible~\cite{petrietal02}, these models still suffer from the lack
of a self-consistent solution for the pulsar magnetosphere.
Nevertheless, recent observations of the polarisation of the optical
pulses from the Crab~\cite{kanbachetal03} motivated detailed
comparative studies of the emission predicted by the polar cap, the
outer gap and the two-pole caustics models~\cite{dykshardingrudak04}.
In all of these models, the radiation is produced within the light
cylinder. However, the pulse profile is determined by the assumed
geometry of the magnetic field and the location of the gaps. None of
these models is able to fit the optical polarisation properties of the
Crab pulsar.

An alternative site for the production of pulsed radiation has been
investigated~\cite{kirkskjaeraasengallant02}, based on the idea of a
striped pulsar wind, originally introduced by~\cite{coroniti90}
and~\cite{michel94}.  Emission from the striped wind originates
outside the light cylinder and relativistic beaming effects are
responsible for the phase coherence of the synchrotron radiation. A
strength of this model is that the geometry of the magnetic field,
which is the key property determining the polarisation properties of
the emission, is relatively well-known.

In this work, we use an explicit asymptotic solution for the
large-scale field structure related to the oblique split monopole and
valid for the case of an ultra-relativistic plasma~\cite{bogovalov99}.
This is combined with a crude model for the emissivity of the striped
wind and of the magnetic field within the dissipating stripes
themselves. We calculate the polarisation properties of the
high-energy pulsed emission and compare our results with optical
observations of the Crab pulsar.

\section{Stokes parameters}

\begin{figure}
  \centerline{\psfig{file=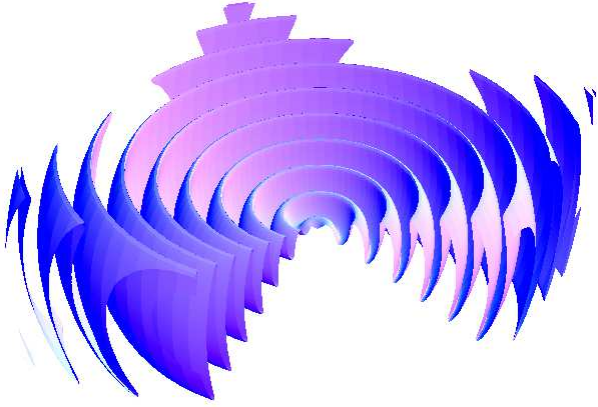,width=8cm,clip=} }
  \caption{3D structure of the current sheet in the striped wind. 
    The rotating neutron star is located at the origin of the
    coordinate system.
    \label{fig:wind}}
\end{figure}
Our magnetic field model is based on the asymptotic solution of the
split monopole for the oblique rotator, valid for $r\gg r_{\rm L}$,
and modified to take account of a finite width of the current sheet.
We add a small meridional component in this sheet in order to prevent
the magnetic field from becoming identically zero.  The 3-dimensional
geometry of the current sheet is shown in figure~\ref{fig:wind}.  In
spherical polar coordinates $(r,\theta,\varphi)$ centered on the star
and with axis along the rotation axis, the radial field is small and
neglected, $ B_r \sim B_{\rm L} \, r_{\rm L}^2 / r^2 $, while the other
components are:
\begin{eqnarray}
  \label{eq:BFBT}
  B_\theta & = & B_{\rm L} \, \frac{r_{\rm L}}{r} \, b_{1,2} \,
  \eta_{\theta}(\Delta_\theta, r,\theta,\varphi,t) \\
  B_\varphi & = & B_{\rm L} \, \frac{r_{\rm L}}{r} \, 
  \eta_\varphi(\Delta_\varphi,r,\theta,\varphi,t) \\
  \eta_\varphi(\Delta_\varphi, r,\theta,\varphi,t) & = &
  \tanh\left[\Delta_\varphi\, \left( \cos\theta \, \cos\alpha + \right. \right. \\
  & & \left. \left. \sin\theta \, \sin\alpha \, \cos \left\{\varphi - \Omega_* \,
        \left( t - \frac{r}{v} \right) \right\} \right) \right]
  \nonumber\\
  \eta_\theta(\Delta_\theta, r,\theta,\varphi,t) & = &
  \frac{1}{\Delta_\theta} \, \frac{\partial \eta_\varphi(\Delta_\theta,
    r,\theta,\varphi,t)}{\partial \varphi}
  \label{magneticfield}
\end{eqnarray}
Here, $B_{\rm L}$ is a fiducial magnetic field strength, $v$ is the
(radial) speed of the wind, $\Omega_*=c/r_{\rm L}$ is the angular
velocity of the pulsar, with $r_{\rm L}$ the radius of the light
cylinder, $\alpha$ is the angle between the magnetic and rotation
axes, $b_{1,2}$ are parameters controlling the magnitude of the
meridional field in the two current sheets present in one wavelength,
and $\Delta_{\theta,\varphi}$ are parameters quantifying the sheet
thickness. The functional form of $B_\varphi$ is motivated by exact
equilibria of the planar relativistic current sheet
\cite{kirkskjaeraasen03}.  However, in these equilibria the $B_\theta$
component, which has an important influence on the polarisation
sweeps, is arbitrary.  The $B_\theta$ we adopt corresponds to a small
circularly polarised component of the pulsar wind wave, such as is
expected if the sheets are formed by the migration of particles within
the wave, as described qualitatively by \cite{michel71}.
 
For the particle distribution, we adopt an isotropic electron/positron
distribution given by 
\begin{equation}
  \label{eq:Densite}
  N(E,\vec{p},\vec{r},t) = K(\vec{r},t) \, E^{-p}  
\end{equation}
where $K(\vec{r},t)$ is related to the number density of emitting
particles.  The radial motion of the wind imposes an overall $1/r^2$
dependence on this quantity, which is further modulated because the
energization occurs primarily in the current sheet. The precise value
in each sheet is chosen to fit the observed intensity of each
sub-pulse.  In addition, a small dc component is added, giving the
off-pulse intensity.  For the emissivity, we use the standard
expressions for incoherent synchrotron radiation of ultra-relativistic
particles.  We assume the emission commences when the wind crosses the
surface $r=r_0\gg r_{\rm L}$.

The calculation of the Stokes parameters as measured in the observer
frame involves simply integrating the emissivity over the wind.  For
an observer, at time~$t_{\rm obs}$, they are given by the following
integrals:
\begin{eqnarray}
  \label{eq:StokesParameters}
  \left\{
    \begin{array}{c}
      I_\omega \\
      Q_\omega \\
      U_\omega
    \end{array}
  \right\} (t_{\rm obs}) & = & \!\!\! \int_{r_0}^{+\infty} \!\!\!\!\! 
  \int_0^{\pi} \!\!\! \int_0^{2\,\pi} \!\!\! \!\!\! s_0(\vec{r},t_{\rm ret})  
  \left\{
    \begin{array}{c}
      \frac{p+7/3}{p+1} \\
      \cos \, (2\,\tilde{\chi}) \\
      \sin \, (2\,\tilde{\chi})
    \end{array}
  \right\} d^3\vec{r}
\end{eqnarray}
where the retarded time is given by $t_{\rm ret} = t_{\rm obs} +
\vec{n}\cdot\vec{r}/c$ and $\vec{n}$ is a unit vector along the line
of sight from the pulsar to the observer.  In this approximation the
circular polarisation vanishes: $V=0$. The function~$s_0$ is defined
by:
\begin{eqnarray}
  \label{eq:ParaStokes}
  s_0(\vec{r},t) & = & \kappa \, K(\vec{r},t) \,
  \frac{\mathcal{D}^{\frac{p+3}{2}}}{\omega^{\frac{p-1}{2}}}
  \, \left( \frac{B}{\Gamma} \, \sqrt{ 1 - ( \mathcal{D} \, \vec{n} \cdot \vec{b} )^2 } 
  \right)^{\frac{p+1}{2}}
\end{eqnarray}
where $\omega$ is the angular frequency of the emitted radiation, and
$\kappa$ is a constant factor that depends only on the nature of the
radiating particles~(charge $q$ and mass $m$) and the power law index
$p$ of their distribution:
\begin{eqnarray}
  \kappa & = & \frac{\sqrt{3}}{2\,\pi} \, \frac{1}{4} \, \Gamma_{\rm Eu} 
  \left( \frac{3\,p+7}{12}\right) \, 
  \Gamma_{\rm Eu}\left(\frac{3\,p-1}{12}\right) \, \nonumber \\ & &
  \frac{|q|^3}{4\,\pi\,\varepsilon_0\,m\,c} 
  \, \left( \frac{3\,|q|}{m^3\,c^4} \right)^{\frac{p-1}{2}} 
\end{eqnarray}
with $\Gamma_{\rm Eu}$ the Euler gamma function and $\mathcal{D}$ the
Doppler boosting factor $\mathcal{D} = 1/\Gamma \, ( 1 - \vec{\beta}
\cdot \vec{n} )$. The direction of the local magnetic field in the
observer's frame is given by the unit vector $\vec{b}$ and the
simplifying assumption has been made that this field has no component
in the direction of the plasma velocity: $\vec{b}\cdot\vec\beta=0$.
(In this case the magnetic field transformation from the rest
frame~$\vec{B}'$ to the observer frame~$\vec{B}$ is just $\vec{B}' =
\vec{B}/\Gamma$ and, thus, its direction remains unchanged.)  The
angle~$\tilde{\chi}$ measures the inclination of the {\it local\/}
electric field with respect to the projection of the pulsar's rotation
axis on the plane of the sky as seen in the observer's frame. The
degree of linear polarisation is defined by 
\begin{equation}
  \label{eq:Pi}
  \Pi = \frac{\sqrt{Q^2 + U^2}}{I}  
\end{equation}
The corresponding polarisation angle, defined as the position angle
between the electric field vector at the observer and the projection
of the pulsar's rotation axis on the plane of the sky is
\begin{equation}
  \label{eq:Chi}
  \chi = \frac{1}{2} \, \arctan \left( \frac{U}{Q} \right)
\end{equation}

\section{Results}

\begin{figure}
  \centerline{\psfig{file=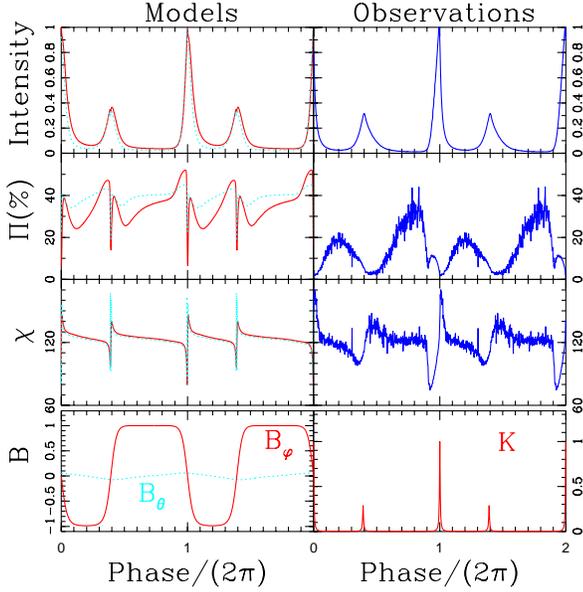,width=8.cm,clip=} }
  \caption{Light curve of intensity, degree of polarisation and 
    position angle of the pulsed synchrotron emission obtained for our
    model and measurements of these quantities for the Crab pulsar.
    Models with Lorentz factor~$\Gamma=20$ (solid red) and $50$
    (dotted cyan) are shown. The bottom panels show the dependence on
    phase of the assumed magnetic field components and the particle
    density in the comoving frame.
    \label{fig:polar}}
\end{figure}

The upper left panel of Fig.~\ref{fig:polar} shows the intensity
(Stokes parameter $I$) computed using our smoothed profile with
$\Delta_\theta=1$, $\Delta_\varphi=5$, $b_1=0.1$ and $b_2=0.08$ for
each sub-pulse. The electron density is 
\begin{equation}
  \label{eq:KK}
  K = \frac{1}{ r^2 \, ( 1 - 0.6\,\eta_\theta) } \, 
  \left[ \left( \frac{r_{\rm L} \, B_{\rm L} }{ r \, B } \right)^{(p+1)/2} + \varepsilon - 1 \right]
\end{equation}
where the parameter $\varepsilon=0.05$ sets the minimum electron
density between the current sheets (in normalized units). The
denominator~$(1-0.6\,\eta_\theta)$ introduces an asymmetry in the
relative pulse peak intensity. The variation of the magnetic field and
the particle density along the line of sight, are shown in the bottom
panels of Fig.~\ref{fig:polar}.
  
The upper panels of Fig.~\ref{fig:polar} show the results of our
computations on the left and the corresponding observed quantities
\cite{kanbachetal03} on the right. Comparison with the upper
right-hand panel shows that the model reproduces the observed pulse
profile quite accurately. In this example, we adopt an obliquity
$\alpha=60\degr$ and an inclination of the rotation axis to the line
of sight $\xi = \arccos( \vec{n} \cdot \vec{\Omega}_* /
|\vec{\Omega}_*| ) = 60 \degr$ and set the radius at which emission
switches on to be $r_0 = 30r_{\rm L}$. The position angle of the
projection of the pulsar's rotation axis is set to $124\degr$
\cite{ngromani04}. The electron power law index $p=2$, as suggested by
the relatively flat spectrum displayed by the pulsed emission between
optical and gamma-ray frequencies \cite{kanbach98}.  Results are shown
for two values of the Lorentz factor of the wind: $\Gamma=20$ (solid
line) and $\Gamma=50$ (dotted line).  For convenience, the maximum
intensity is normalized to unity. The timescale is expressed in terms
of pulse phase, $0$ corresponding to the initial time~$t=0$ and $1$ to
a full revolution of the neutron star and thus to one period~$T_* =
2\pi / \Omega_*$.

The degree of polarisation is shown in the two middle panels of
Fig.~\ref{fig:polar}. According to our computations (left-hand panel) this
displays a steady rise in the initial off-pulse phase, that steepens
rapidly as the first pulse arrives.  During the pulse phase itself,
the polarisation shrinks down to about 10\%. Theoretically the maximum
possible degree of polarisation is closely related to the index~$p$ of
the particle spectrum.  In the most favorable case of a uniform
magnetic field, it is given by 
\begin{equation}
  \label{eq:Pmax}
  \Pi_{\rm max} = \frac{p+1}{p+7/3}
  \end{equation}
However, in the curved magnetic field lines of the wind, contributions
of electrons from different regions have different polarisation
angles. Consequently, they depolarise the overall result when
superposed. We therefore expect a degree of polarisation that is at
most~$\Pi_{\rm max}$. For the example shown in figure~\ref{fig:polar},
$\Pi_{\rm max}(p=2) = 69.2$\%, well above the computed value, which
peaks at~$52$\%.

The lower panels of Fig.~\ref{fig:polar} show the polarisation angle,
measured against celestial North and increasing from North to East.
Our model predicts this angle relative to the projection of the
rotation axis of the neutron star on the sky, which we take to lie at
a position angle of $124\degr$, following the analysis of
\cite{ngromani04}.  In the off-pulse stage, the electric vector of our
model predictions lies almost exactly in this direction, since it is
fixed by the orientation of the dominant toroidal
component~$B_\varphi$ of the magnetic field.  In the rising phase of
the first pulse, $B_\varphi$ decreases, whereas $B_\theta$ increases,
causing the polarisation angle to rotate from its off pulse value by
about $50\degr$, for the chosen parameters.  However, for a weaker
$B_\theta$ contribution, as in the second pulse, the swing decreases.
This effect can also be caused by a relatively large beaming angle,
(i.e., low Lorentz factor wind).  The basic reason is that
contributions from particles well away from the sheet center are then
mixed into the pulse, partially canceling the contribution of the
particles in the center of the current sheet, which favor $\chi = 124
\degr \pm 90 \degr$, and enforcing~$\chi = 124\degr$.  On the other
hand, a very high value of the Lorentz factor or large values of
$b_{1,2}$ reduce the off-center contribution, leading, ultimately, to
the maximum possible $90\degr$ sweep between off-pulse
($B_\varphi$-dominated) and center-pulse ($B_\theta$-dominated)
polarisations, followed by another $90\degr$ sweep in the same sense
when returning to the off-pulse. Thus, in general, in the middle of
each pulse, the polarisation angle is either nearly parallel to the
projection of the rotation axis, or nearly perpendicular to it,
depending on the strength of~$B_\theta$ and on~$\Gamma$.  This
interpretation is confirmed by computations with $\Gamma=50$ that show
a larger sweep, as shown in Fig.~\ref{fig:polar}.

The optical polarisation measurements suggest that in the centre of
the pulses the position angle is close to $124\degr$.  In the
declining phase of both pulses, the angle reaches a maximum before
returning to the off-pulse orientation.  Note that in both cases the
swing starts in the same direction, (counterclockwise in
figure~\ref{fig:polar}). This is determined by the rotational behavior of
the $B_\theta$ component, implying that this changes sign between
adjacent sheets, as in Eq.~(\ref{magneticfield}).  The observed
off-pulse position angle is closely aligned with the projection of the
rotation axis of the pulsar, in accordance with the model predictions.

In addition to models aimed at providing a framework for the
interpretation of the emission of the Crab pulsar, we have performed
several calculations with different Lorentz factors~$\Gamma$,
injection spectrum of relativistic electrons~$p$ and inclinations of
the line of sight~$\xi$. The general characteristics of the results
are: For low Lorentz factors, independent of $p$ and $\xi$, the
relativistic beaming becomes weaker and the pulsed emission is less
pronounced, because the observer receives radiation from almost the
entire wind. For instance, taking~$\Gamma=2$ and $p=2$ or $3$, the
average degree of polarisation does not exceed 20~\% and the swing in
the polarisation angle is less than $30\degr$. For high Lorentz
factors $\Gamma\ge50$, the strong beaming effect means that the
observer sees only a small conical fraction of the wind. The width of
the pulses is then closely related to the thickness of the transition
layer. The degree of linear polarisation flattens in the off-pulse
emission while it shows a sharp increase followed by a steep decrease
during the pulses. Due to the very strong beaming effect, only a tiny
part of the wind directed along the line of sight will radiate towards
the observer. In the off-pulse phase, the polarisation angle is then
dictated solely by the $B_\varphi$ component \lq\lq attached\rq\rq\ to
the line of sight, and the degree of linear polarisation remains
almost constant in time.  For very high Lorentz factors, the behavior
of polarisation angle and degree remain similar to those of
Fig.~\ref{fig:polar}, with perfect alignment between polarisation
direction (electric vector) and the projection of the pulsar's
rotation axis on the plane of the sky in the off-pulse phase and two
consecutive polarisation angle sweeps of $90\degr$ in the same sense
during the off-pulse to center-pulse and center-pulse to off-pulse
transitions.  This mirrors the fact that emission comes only from a
narrow cone about the line of sight of half opening angle~$\theta
\approx 1 / \Gamma$.

For given values of~$\Gamma$ and $\xi$, the particle spectral index
$p$ affects only the average degree of polarisation degree but not the
light curve nor the polarisation angle. For example, taking
$\Gamma=10$ and~$\xi=60\degr$, a spectral index of~$p=2$ leads to
an average polarisation of~$\tilde{\Pi}=19.2\%$ whereas for~$p=3$ it
leads to~$\tilde{\Pi}=30.8\%$.

\section{Conclusions}

In the striped wind model, the high energy (infra-red to gamma-ray)
emission of pulsars arises from outside the light cylinder.  It
provides an alternative to the more intensively studied gap models.
They all contain essentially arbitrary assumptions concerning the
configuration of the emission region and the distribution function of
the emitting particles, rendering it difficult to distinguish between
them on the basis of observations. However, the geometry of the
magnetic field, which is the crucial factor determining the
polarisation properties, is constrained in the striped model to be
close to that of the analytic asymptotic solution of the split
monopole.  We have therefore presented detailed computations of the
polarisation properties of the pulses expected in this scenario. These
possess the characteristic property, unique amongst currently
discussed models, that the electric vector of the off-pulse emission
is aligned with the projection of the pulsar's rotation axis on the
plane of the sky.  This is in striking agreement with recent
observations of the Crab pulsar.  In addition the striped wind
scenario naturally incorporates features of the phase-dependent
properties of the polarisation angle, degree of polarisation and
intensity that are also seen in the data.  This underlines the need to
develop the model further, in order to confront high-energy
observations of the Crab and other pulsars.  In particular, the manner
in which magnetic energy is released into particles in the current
sheet remains poorly understood and the link between the asymptotic
magnetic field structure and the pulsar magnetosphere is obscure.

\begin{acknowledgements}
  We thank Gottfried Kanbach for providing us with the OPTIMA data and
  for helpful discussions. This work was supported by a grant from the
  G.I.F., the German-Israeli Foundation for Scientific Research and
  Development.
\end{acknowledgements}



              \clearpage

\end{document}